\documentclass{article}

\usepackage[final, nonatbib]{neurips_2021_ml4ps}

\usepackage[utf8]{inputenc} 
\usepackage[T1]{fontenc}    
\usepackage{hyperref}       
\hypersetup{urlcolor=black,linkcolor=black,citecolor=black}
\usepackage{url}            
\usepackage{booktabs}       
\usepackage{amsfonts}       
\usepackage{nicefrac}       
\usepackage{microtype}      
\usepackage{xcolor}         
\usepackage{wrapfig}
\usepackage{adjustbox}
\usepackage[normalem]{ulem}

\usepackage{graphicx}
\usepackage[numbers,sort&compress]{natbib}
\title{
Probing the Structure of String Theory Vacua with Genetic Algorithms and Reinforcement Learning
}

\author{
    Alex Cole \\
    University of Amsterdam \\
    \texttt{a.e.cole@uva.nl} \\
    \And
    Sven Krippendorf \\
    Arnold Sommerfeld Center for Theoretical Physics \\
    LMU Munich \\
    \texttt{sven.krippendorf@physik.uni-muenchen.de} \\
    \And
    Andreas Schachner \\
    Centre for Mathematical Sciences \\
    University of Cambridge\\
    \texttt{as2673@cam.ac.uk} \\
    \And
    Gary Shiu \\
    University of Wisconsin-Madison \\
    \texttt{shiu@physics.wisc.edu} \\
}

\begin{document}

\maketitle

\begin{abstract}
  Identifying string theory vacua with desired physical properties at low energies requires searching through high-dimensional solution spaces -- collectively referred to as the string landscape. We highlight that this search problem is amenable to reinforcement learning and genetic algorithms. In the context of flux vacua, we are able to reveal novel features (suggesting previously unidentified symmetries) in the string theory solutions required for properties such as the string coupling. In order to identify these features robustly, we combine results from both search methods, which we argue is imperative for reducing sampling bias.
  
\end{abstract}

\section{Motivation}

While the set of solutions arising in string theory -- known as the string landscape \cite{Bousso:2000xa, Susskind:2003kw} -- encompasses a vast number of low energy effective field theories, only a small subset of them are phenomenologically viable.
The landscape's enormity, with as many as $10^{500}$ \cite{Ashok:2003gk,Denef:2004ze} to $10^{272,000}$~\cite{Taylor:2015xtz} states, makes exhaustive searches or random sampling impractical. While in simple string models \emph{generic} solutions may be found easily, practitioners usually seek to construct solutions having special properties such as weak couplings, ensuring theoretical control, or similarity to our universe. These additional criteria impose an auxiliary landscape structure (i.e.\ optimization target) for which constrained systems of equations must be solved.
The existence of a solution with precisely specified properties is not guaranteed -- indeed, the discreteness of UV ingredients generally leads to a rich topological structure of voids and voxels in the landscape's low-energy properties \cite{Denef:2004ze,Cole:2018emh}.
For this problem, the landscape's vastness is exacerbated by computational complexity; finding specific string vacua appears to be NP-hard \cite{Denef:2006ad,Denef:2017cxt,Halverson:2018cio}, though for some toy examples heuristic algorithms may have success \cite{Bao:2017thx}.
On this level, the situation of the string landscape is comparable to spin-glass systems \cite{sherrington1975solvable,barahona1982computational,mezard1987spin} and protein folding \cite{levinthal1969fold,unger1993finding,bryngelson1995funnels,socolich2005evolutionary}.

It remains a top challenge to design efficient optimization methods to search for realistic vacua that may at the same time reveal hidden structure in the landscape. Regarding this latter point, string theorists are interested not only in the construction of realistic vacua but also the statistical distribution surrounding such vacua, which has implications for various dynamics-based proposals for the measure problem \cite{Denef:2017cxt,Bao:2017thx,Khoury:2019yoo}.
In recent years,
stochastic optimization with Genetic Algorithms (GAs)
\cite{Blaback:2013ht,Blaback:2013fca,Abel:2014xta,Cole:2019enn,AbdusSalam:2020ywo} and Reinforcement Learning (RL) \cite{Halverson:2019tkf,Larfors:2020ugo,Krippendorf:2021uxu,Constantin:2021for} have been utilized to search for viable string vacua, outperforming searches based on Metropolis-Hastings. The purpose of the present work is to further develop these approaches and compare their results.
We consider a common search task and apply dimensionality reduction via Principal Component Analysis (PCA) to the results of both algorithms.
We identify two disconnected clusters in the neighborhood of the optimality condition, which are suggestive of a previously unknown symmetry in the distribution of vacua.
We argue that comparing the results of qualitatively different search algorithms is crucial to ensuring the robustness of this structure, effectively reducing sampling bias.

\section{Setup for string theory solutions}

The data structure of the string landscape is characterized by a set of integer inputs, from which a set of continuous parameters are outputted by minimizing a potential energy function. Within the full landscape, we focus on flux vacua where the integer inputs are quantized fluxes $\vec{N}=(N_1,\dots, N_k)$ which are subject to Gauss's law constraints (i.e., $N_i~, i=1, \dots k$ are bounded). The output is a vector of complex-valued fields $\vec{\phi}=(\phi_1, \dots, \phi_m)$ obtained from minimizing a highly non-linear scalar potential $V(\vec{\phi},\vec{N})$ whose functional form is discussed in \cite{Cole:2019enn, Krippendorf:2021uxu}, see also e.g.~\cite{Douglas:2006es, Grana:2005jc} for reviews. The number $k$ of fluxes is related to the number $m$ of scalar fields via $k=4m$. 
In this note, we consider the `conifold limit' in the geometry as defined in Sec.~(5.4) of~\cite{DeWolfe:2004ns} with $k=8$, which fixes the scalar potential $V$.
Our methods generalize to other background geometries, which we plan to study in the future.

The set of minima $\{\vec{\phi}\}$ of $V$ defines the flux landscape. Its elements (flux vacua) have varied physical properties (in the language of GAs, different phenotypes). We focus on the string coupling $g_s$ (parametrizing the way strings interact) and the superpotential $W_0$ (setting the scale of masses).
Our task is solving the inverse problem: given certain physical properties, i.e., values of $(g_s,W_0)$, what is the required input of quantized fluxes $\vec{N}$?
Due to the involved non-linearities, this problem cannot be addressed with closed form solutions and appropriate numerical search strategies have to be applied.
In this work, we compare the efficiency of GAs and RL in solving this inverse problem, thereby aiming to address the following questions:
\begin{enumerate}
    \item Can RL and stochastic optimization such as GAs efficiently search for desirable vacua?
    \item Can these methods discover structure in the landscape?
    \item How is it possible to understand such structure?
\end{enumerate}

\section{The search algorithms}

We apply a generalized version of the GA described in \cite{Cole:2019enn} for generating flux vacua which consists of mainly three steps. We begin with a set of flux vectors $\lbrace\vec{N}_{1},\ldots,\vec{N}_{p}\rbrace$, called a population of size $p$, with $\vec{N}_{i}\in [-30,30]^8$. The GA selects individuals based on a fitness function, and individuals are in turn used to construct new flux vectors $\vec{N}_{l}$ by performing crossover operations acting on pairs of vectors $(\vec{N}_{i},\vec{N}_{j})$. The third and arguably most important step concerns mutation, which alters the vectors $\vec{N}_{l}$ by some randomized procedure. Lastly, one defines a novel population and repeats the above process.
Each individual step involves a choice of operators, see e.g.~\cite{davis1991handbook,Ruehle:2020jrk} for a comprehensive list.
In our implementation, we utilize eight selection, nine crossover and nine mutation operators which, together with mutation, cloning and survival\footnote{Generically, it turns out to be useful to carry over a certain number of the fittest individuals (cloning) or replace the least fit individuals by fitter ones (survival of the fittest).} rate, amounts to a total of $29$ hyperparameters. They are determined for each task separately through sample runs for mini-batches and randomly chosen hyperparameters by extracting their correlations with the mean average distance of the final output to the optimal solution.
This is a novel and powerful approach to optimizing hyperparameters for GAs that is easily generalized to other applications.

We compare the GA with the A3C RL implementation for flux vacua presented in~\cite{Krippendorf:2021uxu} where it was found that A3C resulted in the best performance across several different RL algorithms. Starting from a random initial configuration $\vec{N}_{i}\in [-30,30]^8$ the RL agent can increase or decrease in a single step one of the $k$ entries by one, corresponding to a $16$ dimensional action space. To identify models, the RL agent takes up to $500$ steps or is reset. The policy network, consisting of four hidden dense layers, is optimized based on a reward function which balances the string theory consistency conditions (e.g. Gauss's law constraint) and the phenomenological requirements associated to the scalar potential, the string coupling, and the superpotential.

\section{Results}

\begin{figure}[t!]
  \centering
 \includegraphics[scale=0.2075]{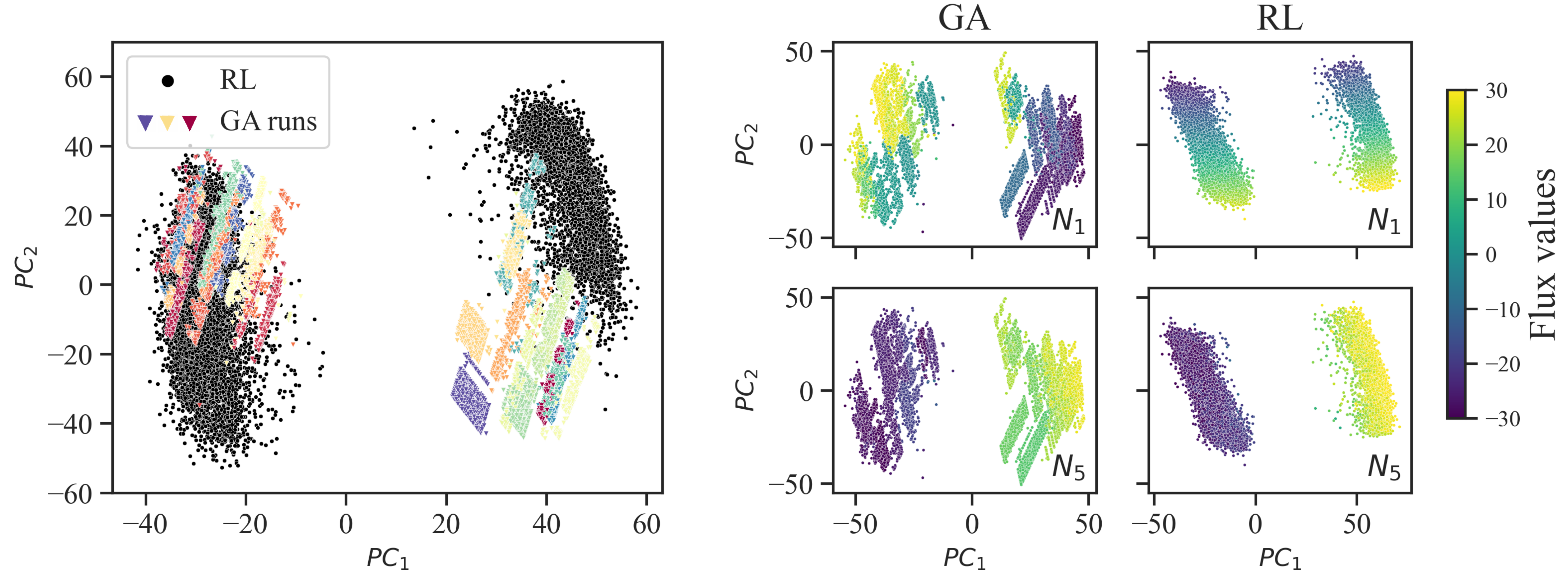}
  \caption{{\it Left:} Cluster structure in dimensionally reduced flux samples for RL and $25$ GA runs (PCA on all samples of GA and RL). The colors indicate individual GA runs. {\it Right:} Dependence on flux (input) values ($N_3$ and $N_5$ respectively) in relation to principal components for a PCA fit of the individual output of GA and RL. 
  }\label{fig:PCAAll}
\end{figure}

For illustrative purposes we chose a target value of $|W_0|=50,000$ which guarantees that many consistent flux vectors can be identified within reasonable effort.
Both RL and GA approaches are found to be extremely efficient in identifying distinct string vacua with desirable properties, with RL producing for an arbitrary starting point an acceptable solution in $193$ steps on average and respectively final GA populations display distinct and desired properties across the population after generating $20$ new populations. For our GA samples it is most efficient to work with population sizes of $500$ using $25$ runs. The optimized choice of hyperparameters for this particular task led to a non-trivial combination of two selection, three crossover and four mutation operators.\footnote{To be more precise, we used tournament and rank selection, random $k$-point, uniform and whole arithmetic recombination crossover as well as random, random $k$-point, insertion and displacement mutation.} Our A3C agent was trained for $15,000,000$ episodes where the agent can take up to $500$ steps before being reset to a random starting point. In both cases we utilize $10,000$ samples for our analysis satisfying $|W_0|=50,000\pm 1,000$.

Being able to efficiently generate samples, we now report on how we can determine and analyze the structure of the set of 8-dimensional flux vectors. This is imperative to develop an understanding of the string landscape and to extract constraints for UV parameters, here our flux vectors, associated with phenomenological properties, i.e.,~$|W_0|=50,000$ for our search. To this end, we perform a PCA on the combined GA and RL samples which is shown on the left of Fig.~\ref{fig:PCAAll}. Uniformly across GA and RL samples we observe two clusters which provides a non-trivial cross-check of this structure. This is remarkable as the sampling dynamics associated to GA and RL are different and we hence consider these features as less likely to be artifact of our sampling procedure. We checked that equivalent structures emerge when performing t-sne \cite{van2008visualizing}. We note that the samples generated from outputs of different GA runs show clear variations. They can be traced back to the way GA operations produce new flux vectors.

The presence of two clusters with uniformly distributed phenomenological properties (superpotential $|W_0|$ and string coupling $g_s$) suggests the emergence of a discrete symmetry whose exact understanding provides an exciting future direction.
Analytically the structure within the clusters can be characterized with scalings of flux components as shown on the right of Fig.~\ref{fig:PCAAll}~where a PCA was performed on the GA and RL samples individually.
While the results are largely equal for $N_5$,
the scaling with respect to $N_1$ highlights distinguishing features between the GA and RL samples.
We find that this structure on the "consistent" flux vectors highly depends on the imposed phenomenological properties, i.e.,~when sampling different superpotential values the structure changes.

\begin{figure}[t!]
  \centering
  \includegraphics[scale=0.6]{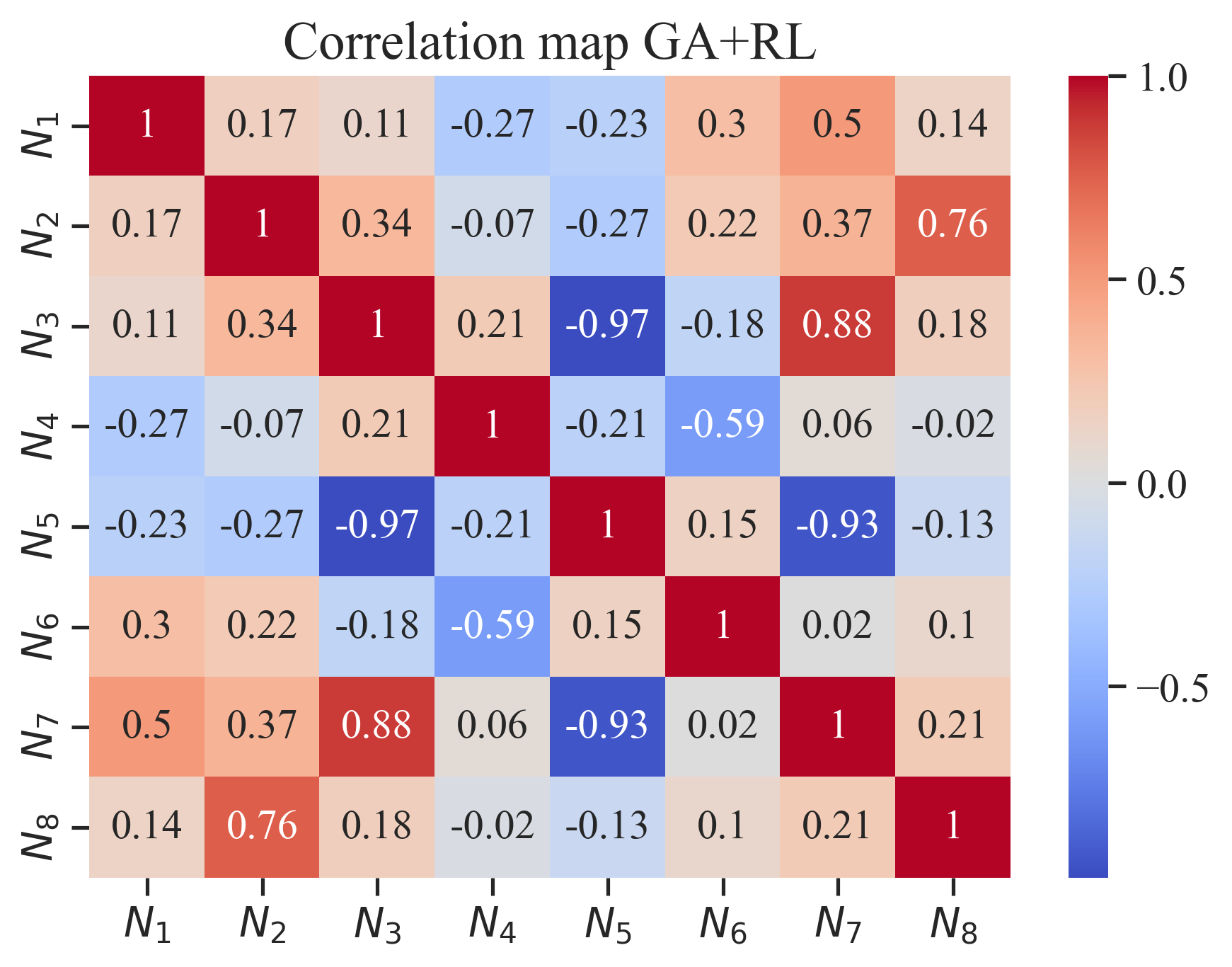}
  \caption{Correlation of flux parameters for vacua with $|W_0|=50,000\pm 1,000$. We use both GA and RL samples. 
  }\label{fig:corr}
\end{figure}

For the PCA analysis we observe the variance ratios for the respective PCA components PC$_i$ listed in Tab.~\ref{tab:PCA}. As this low-variance of the higher PCA components suggests, the sample is effectively lower dimensional. This can be understood further by checking the correlations among the flux vacua. Similar to \cite{Krippendorf:2021uxu}, we compare them for both search strategies which is capable of highlighting universal structures specific to a given class of vacua.
\begin{wraptable}{l}{0.45\linewidth}
\vspace*{-0.25cm}
\setlength{\tabcolsep}{2pt}
\caption{Variance of components in PCA.}
  \label{tab:PCA}
  \centering
  \begin{tabular}{l||c|c|c|c|c}
    \toprule
    Data& PC$_1$ & PC$_2$ & PC$_3$ & PC$_4$ & PC$_5$ -- PC$_8$    \\[0.2em]
   \hline
   \hline
   &&&&\\[-0.8em]
   GA+RL & 0.52& 0.22& 0.13& 0.1&  <0.03 \\[0.2em]
   \hline
   &&&&\\[-0.8em]
   GA & 0.60 & 0.19& 0.13& 0.05 &<0.03   \\[0.2em]
   \hline
   &&&&\\[-0.8em]
    RL &0.68 &0.13 &0.09& 0.08 &<0.02 \\[0.2em]
    \bottomrule
  \end{tabular}
  \vspace*{-0.15cm}
\end{wraptable}
Combining samples from both search algorithms before computing correlations allows us to reduce sampling bias. In Fig.~\ref{fig:corr}, we show the combined correlation maps for the two methods. While certain aspects agree in the individual GA and RL correlations, variations suggest that we ought to consider the combined data set. In particular, the combined correlation maps has strong correlations which are also shared with the individual correlation maps from GA and RL. This includes the correlations of $(N_2 ,N_8)$, $(N_3 ,N_7)$, $(N_5 ,N_7)$ and $(N_3 ,N_5)$,
but, in both GA and RL, we also observe different correlations not present in the respective other sample.
We note that the flux pairs $(N_2 ,N_8)$ and $(N_3 ,N_5)$ appear as products in the D3-charge contribution to the tadpole.
In addition, the dominant PC directions tell us how to move in flux space to find flux vectors with similar physical properties, providing a new direction on studying the statistical properties of this landscape in a dimensionally reduced subspace.

\section{Perspective}
We have demonstrated that by utilizing both RL and GA, it is possible to identify novel structures in the string landscape. To reduce sampling bias and identify these structures robustly, we found it more beneficial to use multiple search strategies. At this stage we cannot single out one method as being preferred. Overall, it should be stressed that the performance is subject to hyperparameters. Here we added and optimized hyperparameter searches in the GA analysis extending previous studies in \cite{Cole:2019enn} and used RL hyperparameters which emerged from a hyperparameter search in~\cite{Krippendorf:2021uxu}.

Optimization becomes very difficult when the reward structure is sparse. We have considered an example where the optimization task is feasible. For model building purposes, it is interesting to consider questions of fine-tuning and rare solutions.
For instance, certain string theory models (e.g.\ \cite{Kachru:2003aw}) assume the existence of rare vacua with $|W_0|\ll 1$. It will be interesting to compare GA and RL strategies to those developed by human string phenomenologists in the search for these vacua (e.g.~\cite{Demirtas:2019sip}). 
It is important to stress that our sampling techniques in this example clearly demonstrate that lower dimensional subsets of the landscape are no obstacle, even though they generally populate a measure zero subset of the state space. The true challenge is presented by the local structure near the global optimum which is generally unknown a priori. We plan to utilize our current methods to explore the local structure near rare string theory solutions obtained in the literature.

Revealing the structure behind and constraints on string theory solutions has a significant potential for our understanding of string theory. Potentially, such constraints could rule out phenomenologically viable combinations of low-energy parameters when no vacua can be constructed in such a region. Hints of such constraints are clearly visible in the current analysis. However, our analysis also shows that -- at this stage -- numerical samples have to be considered carefully, as we observe variations across different methods (cf.~Fig.~\ref{fig:PCAAll}). 
When correlations can be identified robustly, they can be used to efficiently initialize sampling strategies (e.g.\ proposal densities) for vacua with shared properties.
Explicitly constructing the symmetries identified by search algorithms presents another interesting avenue of future work.

Finally, we look forward to investigating how these early results can be transferred to other more complex backgrounds and different phenomenological requirements.

\begin{ack}

We thank the authors and maintainers of
\texttt{Jupyter} \cite{kluyver2016jupyter},
\texttt{Matplotlib} \cite{hunter2007matplotlib},
\texttt{NumPy} \cite{van2011numpy},
\texttt{Seaborn} \cite{waskom2021seaborn},
\texttt{Pandas} \cite{mckinney2011pandas},
\texttt{Openai gym} \cite{brockman2016openai},
\texttt{Scikit-learn} \cite{pedregosa2011scikit} and
\texttt{Chainer-rl} \cite{tokui2015chainer}.
SK would like to thank R.~Kroepsch and M.~Syvaeri for helpful discussions.
AS acknowledges support by the German Academic Scholarship Foundation and by DAMTP through an STFC studentship.
GS is supported in part by the DOE grant DE-SC0017647.

After completion of our work, \cite{Abel:2021rrj} appeared on the arXiv which compares GAs and RL on a different type of search question in string theory, i.e.,~identifying the correct matter content resembling the Standard Model in the context of the heterotic string.

\end{ack}

\bibliography{Reference.bib}

\end{document}